\documentclass[cameraready]{Interspeech}
\title{RIVET: Robust Idempotent Voice Attribute Editing }

\author[affiliation={1}]{Dareen}{Alharthi}
\author[affiliation={1}]{Bhuvan}{Koduru}
\author[affiliation={1}]{Rita}{Singh}
\author[affiliation={1}]{Bhiksha}{Raj}

\address{
    $^1$ Carnegie Mellon University, Pittsburgh, PA, USA
}

\email{dalharth@cs.cmu.edu, bkoduru@cs.cmu.edu, rsingh@cs.cmu.edu, bhiksha@cs.cmu.edu}

\usepackage{comment}
\usepackage{amsmath}
\usepackage{amssymb}

\begin{document}

\maketitle


\begin{abstract}

% Voice attribute editing models aim to modify characteristics such as age, gender, or accent while preserving speaker identity. In practice, attribute annotations in large scale speech datasets are often noisy or inconsistent. When trained under such supervision, conditional generative models can entangle identity with attributes and produce unstable edits. In this work, we show that idempotency provides an effective mechanism for improving robustness to noisy labels. In our setting, idempotency means that the model output is stable under repeated processing, so re encoding and decoding a generated sample returns to the same latent representation and does not drift. This stability constraint acts as an implicit regularizer, reducing sensitivity to mislabeled examples. We introduce RIVET (Robust Idempotent Voice Attribute Editing), a training framework that incorporates an idempotency objective into conditional voice editing models. We evaluate RIVET on the GLOBE dataset, which contains noisy attribute labels. RIVET improves editing success rates and better preserves speaker identity compared to standard conditional training. These results show that idempotency is not only useful for stable generation, but also an effective way to train robust voice editing models under noisy supervision. 

Voice attribute editing models modify characteristics such as age and gender
while preserving speaker identity. In large-scale speech datasets, however,
attribute annotations are often noisy or inconsistent, which can cause
conditional generative models to produce unstable edits. In this work, we show that idempotency provides an effective mechanism
for improving robustness to noisy labels. An idempotent operator is one
for which repeated application does not change the result, i.e.,
$f(f(x)) = f(x)$. Enforcing this property acts as an implicit
regularizer that reduces sensitivity to mislabeled examples. We introduce RIVET, a training framework that incorporates an
idempotency objective  to improve
robustness to label noise. We evaluate RIVET under controlled label noise and on the GLOBE dataset with naturally noisy annotations. RIVET improves editing success and better preserves speaker identity than standard training, showing that idempotency improves robustness in voice editing models. Code is available online.\footnotemark

\end{abstract}

\footnotetext{
\url{https://github.com/DareenHarthi/rivet}}

\section{Introduction}

Voice attribute editing aims to modify specific characteristics of a speech signal,
such as age, gender, or accent, while preserving the speaker’s underlying identity.
Recent generative models have made significant progress in enabling controllable
speech editing through conditional synthesis and disentangled representations
\cite{anastassiou2024voiceshop, sheng2025voice, ye2024flashspeech}.
In these systems, editing is typically achieved by conditioning the model on
attribute labels or by manipulating dedicated latent variables corresponding to
interpretable factors of variation. However, methods that rely on explicit attribute supervision critically depend on the quality of the labels used during training. In practice, large-scale speech datasets often contain noisy or inconsistent
attribute annotations. Labels may be flipped, ambiguous, weakly defined, or inferred automatically rather
than verified manually  \cite{chen2024imprecise, wang2024globe}.

When trained with noisy supervision, deep generative models may learn incorrect
associations between attributes and acoustic patterns, leading to degraded
conditional generation quality. Prior work has shown that models trained on
noisy labels can memorize spurious correlations and exhibit reduced
generalization performance \cite{frenay2013classification}. In conditional
generative settings such as voice editing, noisy attribute conditioning can
distort the learned conditional mappings, resulting in unstable editing
behavior and entanglement between speaker identity and the specified attribute
\cite{cong2025guiding, na2024label}. 

In the setting of editing, noisy or imprecise training labels directly affect the learned
mapping from an input with a given attribute to the output in which that
attribute has been modified. Repeated application of this learned,
noisy editing process can lead to progressive drift, which drags both the
edited attribute and the underlying speech representation itself away from their
intended values. One way of minimizing the influence of the noise in the label is to minimize this drift: intuitively, a sequence of edits that eventually
restores an attribute to its original value should also return the speech
representation to its original, unedited form. This observation naturally leads to the
concept of \emph{idempotency} \cite{shocher2023idempotent, durasov20243, jensenEnforcing,zaman2025score}.

An idempotent model satisfies the property that once an output lies on the
target data manifold, further applications of the model do not change it.
Formally, for an input $x$, an idempotent function $f$ satisfies $f(f(x)) = f(x)$.
This property encourages stability under repeated generation and prevents
drift during editing. Several recent works have explored enforcing idempotency
in generative models
\cite{shocher2023idempotent, zaman2025score, jensenEnforcing, song2023consistency}.
While some approaches suffer from issues such as blurriness or mode collapse
\cite{shocher2023idempotent}, and others are difficult to scale to more complex
architectures \cite{zaman2025score, jensenEnforcing}, these studies suggest
that idempotency can serve as a powerful constraint for stabilizing
generative models.

In this work, we leverage idempotency as a regularization principle for
attribute-conditioned voice editing under noisy supervision. We introduce RIVET (\emph{Robust Idempotent Voice Attribute Editing}), a training
framework that incorporates an idempotency objective into a conditional voice
editing model, similar in spirit to prior work on idempotent generative models
\cite{shocher2023idempotent}. Instead of enforcing idempotency directly in the
output space, RIVET applies the constraint in the latent representation space,
encouraging the model to map inputs to stable points on the attribute-conditioned
manifold and reducing sensitivity to mislabeled training examples.

We evaluate RIVET on the GLOBE dataset \cite{wang2024globe}, which contains
naturally occurring label noise, and on the EARS dataset \cite{richter2024ears} with controlled
levels of synthetic label noise. Our experiments focus on two attributes,
age and gender. The results show that RIVET improves editing success rates
compared to strong baselines while preserving speaker identity. Although our
evaluation focuses on these two attributes, the principle of enforcing
idempotency is model-agnostic and can be applied to other attributes and editing architectures. Our contributions are summarized as follows:
\begin{itemize}
    \item We demonstrate that idempotency acts as an effective robustness regularizer for attribute-conditioned voice editing under noisy labels.
    \item We introduce RIVET, a simple and general training framework that incorporates idempotent constraints into voice editing models without requiring architectural changes.
    \item We open-source RIVET; to the best of our knowledge, this is the first open-source voice editing framework.
\end{itemize}

The remainder of the paper is organized as follows. 
Section~\ref{sec:related} reviews related work on voice editing and idempotent generative modeling. 
Section~\ref{sec:method} describes the RIVET framework and training objectives. 
Section~\ref{sec:experiments} presents the experimental setup. 
Section~\ref{sec:results} presents the results and discussion. 
Section~\ref{sec:conclusion} concludes with a discussion of implications and future research directions.

\section{Related Work}
\label{sec:related}

\subsection{Voice Editing}

Voice editing aims to modify attributes of a speech signal, such as age, gender, or accent, while preserving speaker identity and linguistic content \cite{anastassiou2024voiceshop, sheng2025voice, lin2024voxgenesis}. A common approach is to learn representations that separate factors of variation so that one attribute can be modified without affecting others. Prior work enforces structured latent spaces \cite{lin2024voxgenesis} or learns attribute-specific directions in embedding space \cite{anastassiou2024voiceshop}, enabling editing through latent manipulation or conditional transformations. However, perfectly disentangling these factors is difficult in practice. Strong constraints may reduce reconstruction quality or model expressiveness \cite{burgess2018understanding, shukor2022semantic}. More recent work instead focuses on preserving identity and content during transformation \cite{chen2023disenbooth, esser2024scaling, rout2024semantic}. Despite these advances, stability remains a challenge: repeated or reversed edits can cause identity drift or artifacts, suggesting that additional constraints are needed to encourage stable editing behavior.

\subsection{Idempotent Models}

Idempotency has recently emerged as a useful property in generative models \cite{shocher2023idempotent, durasov20243, jensenEnforcing,zaman2025score}. A function is idempotent if repeated application does not change the output, meaning that once a sample lies on the data manifold, further applications of the model should leave it unchanged. The Idempotent Generative Network \cite{shocher2023idempotent} introduced explicit losses to encourage stable fixed points in generative models. Later work proposed alternative mechanisms for enforcing idempotent behavior, including distilling idempotent mappings from diffusion model scores \cite{zaman2025score} and using idempotency as a general optimization objective for test-time adaptation in place of auxiliary self-supervised tasks \cite{durasov20243}. Other approaches enforce idempotency through algorithmic updates that progressively move a model toward an idempotent operator during training \cite{jensenEnforcing}.  Most existing work focuses on image generation and assumes clean supervision. In contrast, we apply idempotent constraints to conditional voice editing and show that they can improve stability and robustness under noisy attribute labels.

\subsection{Noisy Labels in Conditional Generation}

Large-scale datasets often contain noisy or unreliable labels due to the difficulty of maintaining high-quality annotations at scale. Prior work has proposed several strategies to mitigate noisy supervision in conditional generative models. One approach incorporates estimates of label reliability or confidence into the conditional input \cite{na2024label, dufour2024don}. Another attempts to correct noisy labels by estimating the underlying clean label distribution or modifying the training objective \cite{cong2025guiding, kaneko2019label}. A third line of work enforces prediction consistency during training. For example, consistency can be encouraged across augmented views of the same sample \cite{englesson2021consistency} or among neighboring samples in representation space \cite{cheng2022class, iscen2022learning}. Such regularization discourages models from fitting incorrect annotations and improves robustness to label noise. In this work, we explore idempotency as a complementary regularization mechanism for improving robustness to noisy labels in conditional voice editing.

\begin{figure}[t]
\centering
\includegraphics[width=\linewidth]{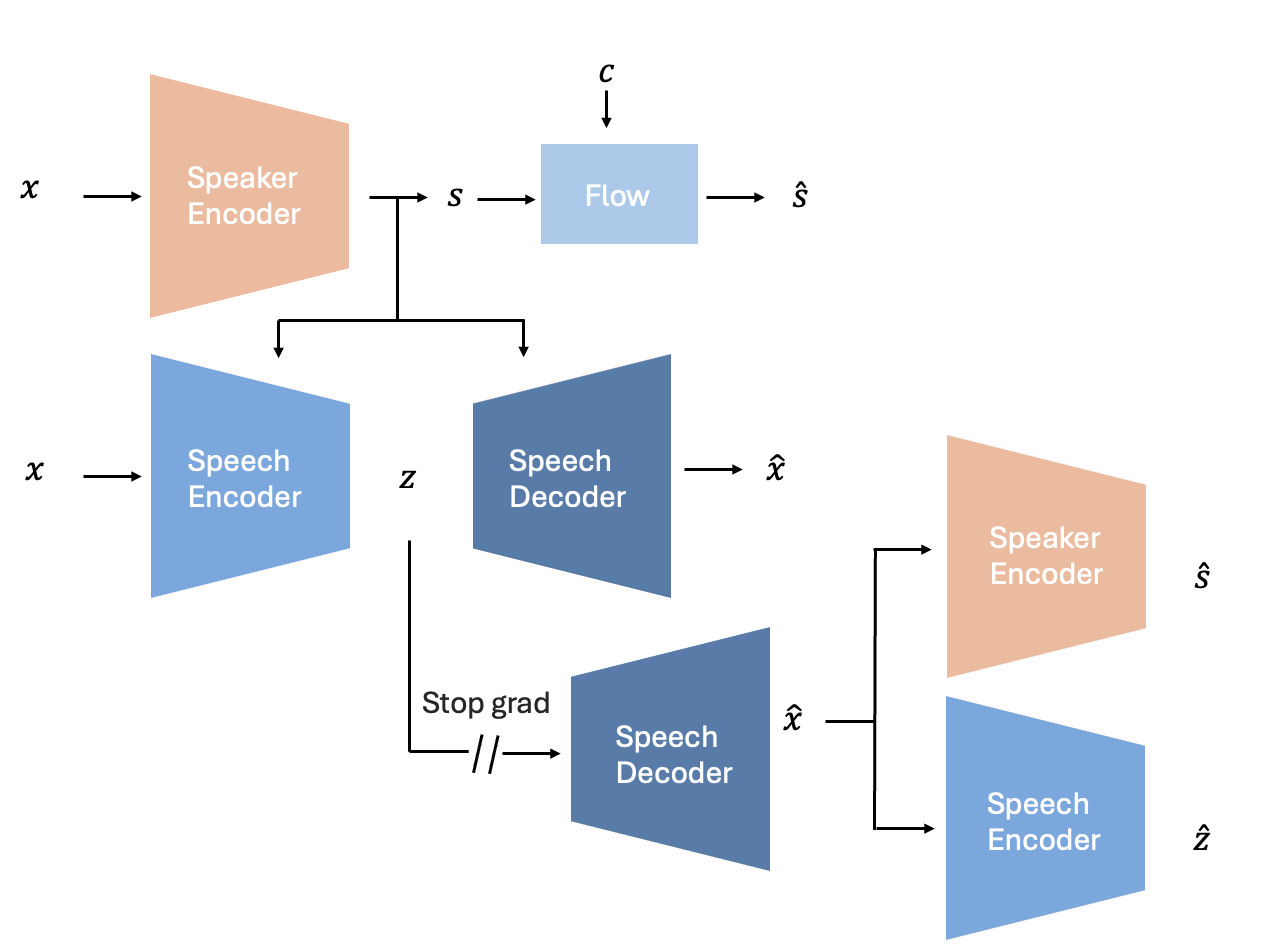}
\caption{
RIVET training framework. The input speech is encoded to obtain speaker and speech representations. The speaker embedding is edited using a conditional flow, and the speech is reconstructed by the generator. The reconstructed speech is then re-encoded to enforce idempotency. A stop-gradient operation is applied before the second decoding step.
}
\label{fig:rivet}
\end{figure}

\begin{figure*}[t]
    \centering
    \includegraphics[width=\textwidth]{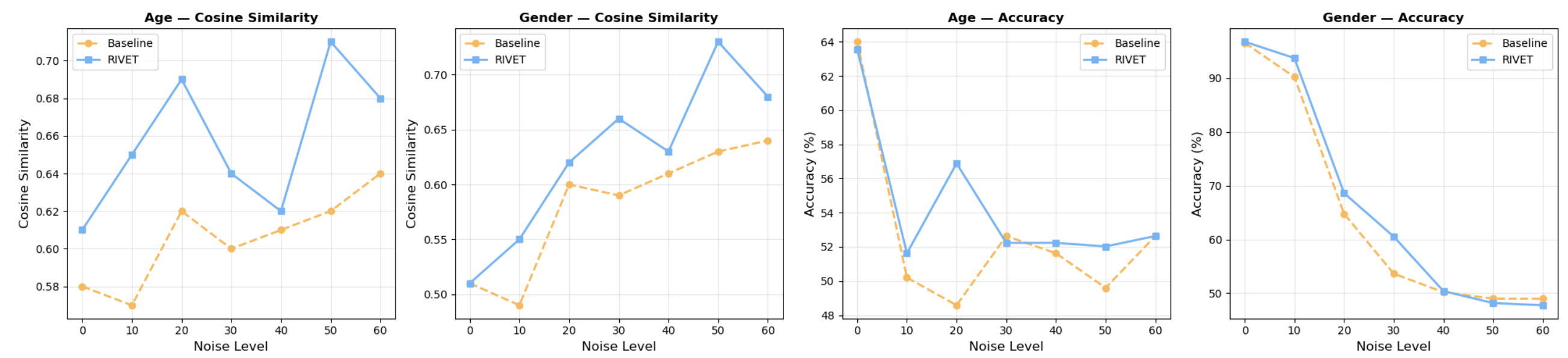}

\caption{
Performance under increasing label noise on the EARS dataset. Models are trained on 7 hours of EARS with increasing noise levels and evaluated on a balanced 1-hour test set. We report cosine similarity between Titanet embeddings of the original and reverted speech (left) and attribute accuracy (right). RIVET maintains higher identity similarity and more stable performance than the baseline as noise increases.
}

    \label{fig:noise_robustness}
\end{figure*}

\section{Method}
\subsection{Idempotent Training }
\label{sec:method}

Let $F(\cdot)$ denote the overall editing model. The model first encodes the
input speech signal $x$ using an encoder $E(\cdot)$ and then reconstructs or
edits the speech using a decoder $D(\cdot)$:

\begin{equation}
F(x) = D(E(x)).
\end{equation}

An operator $F$ is called \emph{idempotent} if repeated application does not
change the result:

\begin{equation}
F(F(x)) = F(x).
\end{equation}

Substituting the encoder–decoder structure gives

\begin{equation}
D(E(D(E(x)))) = D(E(x)).
\end{equation}

A sufficient condition for this relation to hold is that the encoded
representation remains unchanged after reconstruction:

\begin{equation}
E(D(E(x))) = E(x).
\end{equation}

We therefore enforce idempotency by encouraging consistency in the encoded
representation. Let

% \begin{equation}
% z = E(x), \qquad z_{\text{re}} = E(D(E(x))).
% \end{equation}

% The idempotency loss penalizes differences between these representations:

% \begin{equation}
% \mathcal{L}_{\text{idemp}} 
% = 
% \mathbb{E}_{x \sim p_{\mathcal{X}}}
% \left[
% \lVert z - z_{\text{re}} \rVert_2^2
% \right].
% \end{equation}

\begin{equation}
z = E(x), \qquad 
z_{\text{re}} = E(D(E(x))).
\end{equation}

The idempotency loss penalizes differences between these representations:

\begin{equation}
\mathcal{L}_{\text{idemp}}
=
\mathbb{E}_{x \sim p_{\mathcal{X}}}
\left[
\lVert \operatorname{sg}(z) - z_{\text{re}} \rVert_2^2
\right],
\end{equation}

where $\operatorname{sg}(\cdot)$ denotes the stop-gradient operator; it fixes the original latent $z$ as a stable target so gradients update only the re-encoded representation $z_{\text{re}}$, preventing trivial co-adaptation between the two branches. The final training objective combines the standard generative loss with the
idempotency regularizer:

\begin{equation}
\mathcal{L} =
\mathcal{L}_{\text{gen}} +
\lambda \mathcal{L}_{\text{idemp}},
\end{equation}

where $\lambda$ controls the strength of the idempotency constraint.

\subsection{RIVET}

As shown in Figure~\ref{fig:rivet}, RIVET consists of three main components: an ECAPA-TDNN speaker encoder \cite{desplanques2020ecapa}, a conditional normalizing flow for attribute editing \cite{ho2019flow++}, and a VITS-based speech generator \cite{kim2021conditional}. The overall architecture is
similar to VoiceShop \cite{anastassiou2024voiceshop}, but differs in the
training strategy used to enforce idempotency. Given input speech $x$, the ECAPA encoder extracts a fixed-length speaker embedding $\mathbf{e}$ that represents speaker identity. To perform attribute editing, this embedding is transformed using a conditional normalizing flow conditioned on demographic attributes such as age and gender. The flow maps the original embedding $\mathbf{e}$ under the target attribute condition vector $\mathbf{c}$ to produce the edited latent $\mathbf{z}_g$,
\begin{equation}
\mathbf{z}_g, \log |\det J| = f_{\theta}(\mathbf{e}, \mathbf{c}),
\end{equation}
where $J$ denotes the Jacobian of the transformation. The flow is trained with a maximum likelihood objective
\begin{equation}
\label{flow}
\mathcal{L}_{\text{flow}} =
\mathbb{E}
\left[
\frac{1}{2}|\mathbf{z}_g|^2 - \log|\det J|
\right].
\end{equation}

The speaker embedding is used as conditioning information in both the
encoder and decoder of the VITS generator to synthesize the speech signal. 

During training, the generated speech is re-encoded to compute the
idempotency loss described in Section~\ref{sec:method}. As shown in
Figure~\ref{fig:rivet}, the idempotency constraint is applied to both the
speaker encoder (ECAPA-TDNN) and the speech encoder of the VITS generator,
encouraging their representations to remain stable after reconstruction. The ECAPA encoder is additionally trained with attribute classification
objectives for age and gender, denoted by
$\mathcal{L}_{\text{age}}$ and $\mathcal{L}_{\text{gender}}$.
The VITS generator is trained using its standard objectives, including
adversarial, feature matching, mel reconstruction, KL divergence, and
duration losses. The conditional flow is optimized using the maximum
likelihood objective defined in Eq.~\ref{flow}. The overall training objective combines the VITS generator and discriminator
losses, the ECAPA classification losses, the flow likelihood loss, and the
idempotency regularizer applied to both the speaker and speech encoders:

\begin{equation}
\mathcal{L}_{\text{total}} =
\mathcal{L}_{\text{VITS}} +
\lambda_f \mathcal{L}_{\text{flow}} +
\lambda_a \mathcal{L}_{\text{age}} +
\lambda_g \mathcal{L}_{\text{gender}} +
\lambda_i \mathcal{L}_{\text{idemp}},
\end{equation}

where $\mathcal{L}_{\text{VITS}}$ denotes the standard VITS generator and
discriminator losses, and the $\lambda$ terms control the contribution of
each objective. All components are optimized jointly in an end-to-end
training framework. In contrast to VoiceShop
\cite{anastassiou2024voiceshop}, which trains the editing modules and
speaker embeddings separately, RIVET trains all components jointly.

\begin{figure}[t]
\centering
\includegraphics[width=\linewidth]{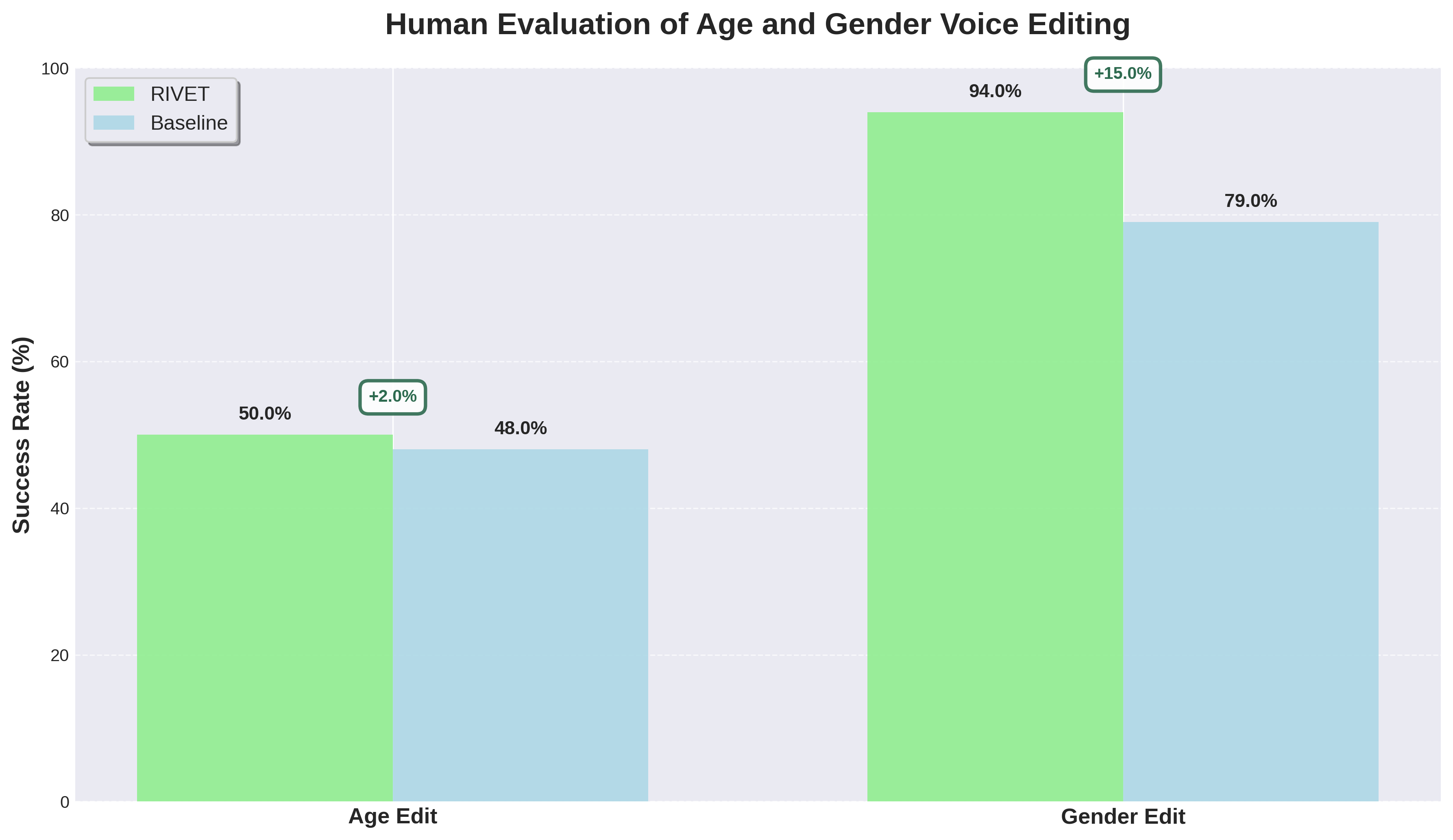}
\caption{
Human evaluation of age and gender voice editing. Each sample was rated by
five annotators with majority voting. RIVET improves
editing success over the baseline.
}
\label{fig:human_eval}
\end{figure}

\section{Experimental Setup}
\label{sec:experiments}

To evaluate the effect of idempotent training, we compare RIVET against a baseline model with identical architecture and training configuration, but without the idempotency constraint. The baseline includes the ECAPA-TDNN speaker encoder, the conditional invertible flow, and the VITS generative backbone, trained jointly using the same objectives. The only difference is that the baseline removes the idempotency regularization term.  For RIVET, we add the idempotency loss to both the speaker embedding space and the speech latent representation. All other training settings, optimization schedules, and hyperparameters are kept identical to ensure a fair comparison. Both models are trained on the GLOBE dataset~\cite{wang2024globe}, which contains approximately 535 hours of English speech from 23{,}519 speakers spanning 164 global accents and a wide age range~\cite{wang2024globe}. The training set is the official GLOBE training split. Quantitative results are reported in Table~\ref{tab:main_results}. To further study robustness to label noise, we conduct controlled experiments using the EARS dataset~\cite{richter2024ears}. EARS is a high-quality speech dataset ~\cite{richter2024ears}. From this dataset, we select only neutral speech samples (excluding emotional recordings). We then construct a training set of approximately 7 hours and a balanced test set of about 1 hour. Using this data, we simulate different levels of label noise by randomly flipping the age and gender labels in the training set. Specifically, we generate training sets with noise levels of 10\%, 20\%, 30\%, 40\%, 50\%, and 60\%. The results are shown in \ref{fig:noise_robustness}.

\section{Results and Discussion}
\label{sec:results}

% Total duration:          31278.22 seconds
%                          521.30 minutes
%                          8.69 hours

% Formatted:               8 hours, 41 minutes, 18 seconds

% \label{sec:results}
% \begin{table}[t]
% \centering
% \small
% \setlength{\tabcolsep}{4pt}
% \caption{
% Cosine similarity between Titanet speaker embeddings of the original speech and the reverted speech,
% and attribute accuracy.
% }
% \label{tab:main_results}
% \begin{tabular}{lcccc}
% \toprule
% \textbf{Method}
% & \multicolumn{2}{c}{\textbf{Cosine}}
% & \multicolumn{2}{c}{\textbf{Accuracy}} \\
% \cmidrule(lr){2-3}
% \cmidrule(lr){4-5}
% & Age & Gender & Age & Gender \\
% \midrule

% \multicolumn{5}{c}{\textbf{GLOBE}} \\
% \midrule
% GT & - & - & 62.8 & 84.9 \\
% Baseline & 0.63 & 0.54 & 39.9 & 77.2 \\
% RIVET & \textbf{0.66} & \textbf{0.55} & \textbf{40.6} & \textbf{85.9} \\

% \midrule
% \multicolumn{5}{c}{\textbf{EARS (OOD)}} \\
% \midrule
% GT & - & - & 99.3 & 99.9 \\
% Baseline & 0.49 & 0.45 & \textbf{32.3} & 74.1 \\
% RIVET & \textbf{0.55} & \textbf{0.48} & 28.7 & \textbf{93.8} \\

% \bottomrule
% \end{tabular}
% \end{table}
\begin{table}[t]
\centering
\small
\setlength{\tabcolsep}{4pt}
\caption{
Cosine similarity between Titanet speaker embeddings of the original speech and the reverted speech,
attribute accuracy, UTMOS, and WER. The models are trained on GLOBE dataset.
}
% \label{tab:main_results}
% \begin{tabular}{lcccccc}
% \toprule
% \textbf{Method}
% & \multicolumn{2}{c}{\textbf{Cosine}}
% & \multicolumn{2}{c}{\textbf{Accuracy}}
% & \textbf{UTMOS}
% & \textbf{WER} \\
% \cmidrule(lr){2-3}
% \cmidrule(lr){4-5}
% & Age & Gender & Age & Gender & Avg & Avg \\
% \midrule

% \multicolumn{7}{c}{\textbf{GLOBE}} \\
% \midrule
% GT & - & - & 62.8 & 84.9 & 3.59 & - \\
% Baseline     & 0.63 & 0.54 & 39.9 & 77.2 & - & - \\
% RIVET        & \textbf{0.66} & \textbf{0.55} & \textbf{40.6} & \textbf{85.9} & - & - \\

% \midrule
% \multicolumn{7}{c}{\textbf{EARS (OOD)}} \\
% \midrule
% GT & - & - & 99.1 & 99.9 & 3.99 & - \\
% Baseline     & 0.49 & 0.45 & \textbf{32.3} & 77.6 & - & - \\
% RIVET        & \textbf{0.55} & \textbf{0.48} & 28.7 & \textbf{93.8} & - & - \\

% \bottomrule
% \end{tabular}
\label{tab:main_results}
\begin{tabular}{lcccccc}
\toprule
\textbf{Method}
& \multicolumn{2}{c}{\textbf{Cosine}}
& \multicolumn{2}{c}{\textbf{Accuracy}}
& \textbf{UTMOS}
& \textbf{WER} \\
\cmidrule(lr){2-3}
\cmidrule(lr){4-5}
& Age & Gender & Age & Gender & Avg & Avg \\
\midrule

\multicolumn{7}{c}{\textbf{GLOBE}} \\
\midrule
GT & - & - & 62.8 & 84.9 & 3.59 & 1.89 \\
Baseline     & 0.63 & 0.54 & 39.9 & 77.2 & 3.17 & \textbf{10.33} \\
RIVET        & \textbf{0.66} & \textbf{0.55} & \textbf{40.6} & \textbf{85.9} & \textbf{3.19} & 10.68 \\

\midrule
\multicolumn{7}{c}{\textbf{EARS (OOD)}} \\
\midrule
GT & - & - & 99.1 & 99.9 & 3.97 & 1.90 \\
Baseline     & 0.49 & 0.45 & \textbf{33.6} & 77.6 & 2.72 & 4.68 \\
RIVET        & \textbf{0.55} & \textbf{0.48} & 30.1 & \textbf{92.7} & \textbf{2.86} & \textbf{4.67} \\

\bottomrule
\end{tabular}
\end{table}

\begin{figure}[t]
\centering
\includegraphics[width=\linewidth]{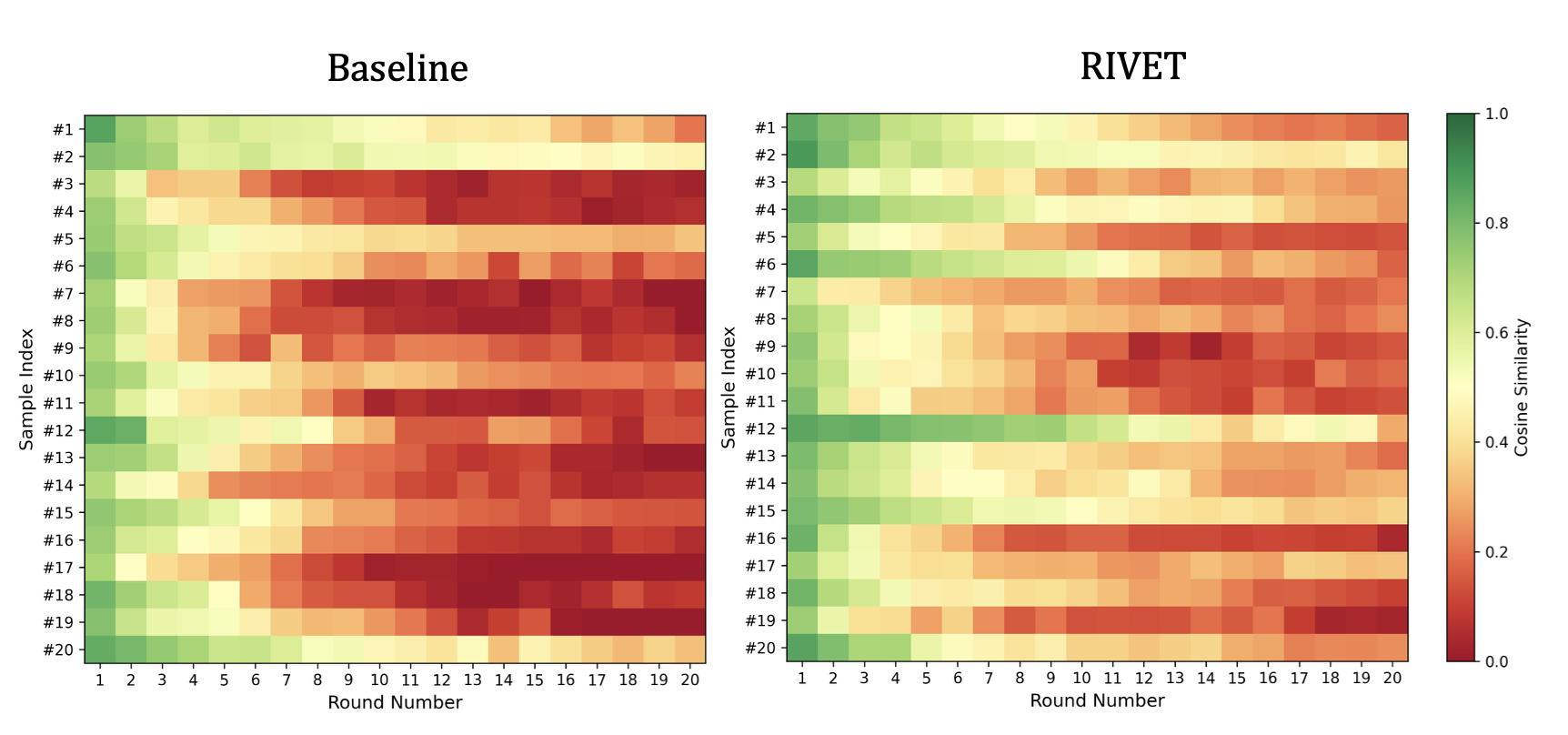}
\caption{
Cosine similarity between Titanet embeddings of the original speech and
reconstructed samples for 20 speakers from the GLOBE test set over 20
reconstruction rounds. Each round uses the output of the previous round as
input. The baseline shows rapid identity drift, while RIVET maintains higher
similarity to the original speaker across iterations.
}
\label{fig:cos_sim}
\end{figure}

\subsection{Evaluation Metrics}

% We evaluate editing performance using three metrics: cosine similarity, attribute accuracy, and human evaluation, UTMOS \cite{saeki2022utmos} to measure naturalness  and WER \cite{radford2023robust} to measure intelligibility. Cosine similarity is computed between Titanet speaker embeddings \cite{koluguri2022titanet} extracted from the original speech and the reverted speech, where the reverted speech is obtained by first editing an attribute (e.g., age or gender) and then reversing the edit back to its original value. This metric measures speaker identity preservation and the quality of the editing operation. Attribute accuracy is computed using classifiers trained independently for age and gender prediction. The classifiers use the same ECAPA-TDNN architecture employed in RIVET and are trained separately on the GLOBE and EARS datasets. These classifiers are used only for evaluation and are not shared with the training of the editing model. Accuracy is
% computed on the edited speech generated by each model. Because GLOBE contains noisy demographic annotations, we train the classifier using a soft-label objective inspired by the EM formulation for learning with imprecise labels \cite{chen2024imprecise}. In GLOBE, age is provided in eight decade-level bins from 10 to 80. For evaluation, we group these bins into three categories—\emph{young} (10–29), \emph{adult} (30–49), and \emph{old} (50+), while gender is treated as a binary classification task. UTMOS and WER are measured on edited speech as well 
We measure speaker identity preservation using cosine similarity between
Titanet speaker embeddings \cite{koluguri2022titanet} extracted from the
original speech and the reverted speech. The reverted speech is obtained by
first editing an attribute (e.g., age or gender) and then reversing the edit
back to its original value. This metric captures both identity preservation
and the stability of the editing process.

Attribute editing success is evaluated using independently trained
classifiers for age and gender prediction. These classifiers use the same
ECAPA-TDNN architecture employed in RIVET but are trained separately on the
GLOBE and EARS datasets and are used only for evaluation. Accuracy is
computed on the edited speech generated by each model. Because GLOBE
contains noisy demographic annotations, the classifiers are trained using a
soft-label objective inspired by the EM formulation for learning with
imprecise labels \cite{chen2024imprecise}. In GLOBE, age is provided in eight
decade-level bins from 10 to 80. For evaluation, we group these bins into
three categories—\emph{young} (10–29), \emph{adult} (30–49), and \emph{old}
(50+), while gender is treated as a binary classification task.
Naturalness and intelligibility are evaluated using UTMOS
\cite{saeki2022utmos} and word error rate (WER) computed using Whisper large-v2 \cite{radford2023robust}. Both metrics are measured on the edited
speech generated by each model.

% Finally, we conduct a perceptual evaluation using Amazon Mechanical Turk,
% where each audio sample is rated by five independent listeners. Annotators
% select the perceived age and gender of the speaker from four options:
% \emph{young female}, \emph{old female}, \emph{young male}, and \emph{old male}.
% For age editing, samples from speakers older than 40 are edited to sound
% younger (approximately 20), while samples from speakers younger than 40 are
% edited to sound older (approximately 50). The final success rate is computed
% using majority voting across listener ratings. The results of the human
% evaluation are shown in Figure~\ref{fig:human_eval}.

Finally, we conduct a perceptual evaluation using Amazon Mechanical Turk,
where each audio sample is rated by five independent listeners. Annotators
select the perceived age and gender from four options: \emph{young female},
\emph{old female}, \emph{young male}, and \emph{old male}. The evaluation
includes 200 samples (25 per category for each model). For age editing,
speakers older than 40 are edited to age 20, while speakers younger than 40
are edited to age 50. Success is computed using majority voting across
listeners. Results are shown in Figure~\ref{fig:human_eval}.
\subsection{Discussion}

% Table~\ref{tab:main_results} reports results on the GLOBE test set and the EARS dataset used as an out-of-distribution (OOD) benchmark. RIVET consistently improves cosine similarity over the baseline, indicating stronger speaker identity preservation. It also improves attribute prediction accuracy on GLOBE, particularly for gender. On EARS, which is not seen during training, RIVET maintains higher cosine similarity and substantially improves gender editing accuracy, while age accuracy slightly decreases. Figure~\ref{fig:cos_sim} illustrates identity stability across repeated reconstruction rounds. The baseline shows progressive identity drift as reconstruction is repeatedly applied, while RIVET maintains higher similarity to the original speaker embedding. Human evaluation in Figure~\ref{fig:human_eval} shows similar trends, with RIVET improving perceived editing success for both attributes, particularly for gender editing. Overall, these results indicate that idempotent training improves identity preservation and stabilizes the editing process.

Table~\ref{tab:main_results} reports results on the GLOBE test set and the
EARS dataset, which we use as an out-of-distribution (OOD) benchmark.
RIVET consistently improves cosine similarity compared to the baseline,
indicating stronger speaker identity preservation. It also improves
attribute prediction accuracy on GLOBE, particularly for gender.
On EARS, which is not seen during training, RIVET maintains higher cosine
similarity and substantially improves gender editing accuracy, while age
accuracy slightly decreases. Both models maintain comparable naturalness
and intelligibility scores. Figure~\ref{fig:cos_sim} illustrates identity stability across repeated
reconstruction rounds. RIVET maintains higher similarity to the original
speaker embedding than the baseline. To further analyze robustness to noisy supervision, we train models on the
EARS dataset with increasing levels of synthetic label noise and evaluate
them on a balanced 1-hour EARS test set. As shown in
Figure~\ref{fig:noise_robustness}, RIVET maintains higher cosine similarity
across noise levels and exhibits more stable behavior as noise increases. Human evaluation results in Figure~\ref{fig:human_eval} show similar
trends, with RIVET improving perceived editing success for both attributes,
particularly for gender editing. Overall, these results show that idempotent training reduces sensitivity to
label noise.

\section{Conclusion}
\label{sec:conclusion}

% In this work, we investigate how idempotency can improve robustness in attribute-conditioned voice editing when training data contains noisy labels. We introduce RIVET, an end-to-end training framework that enforces an idempotency constraint on latent representations. Experiments on GLOBE and the out-of-distribution EARS dataset show improved identity preservation and editing success rate while maintaining comparable perceptual quality. These results suggest that idempotency can serve as an effective regularization mechanism for conditional generative models trained with noisy supervision. We release the RIVET framework to support reproducibility and future research.

% In this work, we studied how idempotency can improve robustness in
% attribute-conditioned voice editing when training data contains noisy labels.
% We introduced RIVET, a unified end-to-end training framework that adds an
% idempotency constraint to the latent representations of a conditional voice
% editing model.  Experiments on the GLOBE and EARS datasets show that idempotent training improves identity preservation and editing
% success rate while maintaining comparable perceptual quality to the baseline. These results
% indicate that idempotency can serve as a practical regularization mechanism
% when attribute labels are imperfect or noisy.  This work takes an initial step toward understanding how idempotent
% objectives improve robustness in conditional generative models. Overall, idempotency
% offers a promising direction for improving stability and robustness
% in voice editing systems.

In this work, we studied how idempotency can improve robustness in
attribute-conditioned voice editing when training data contains noisy labels.
We introduced RIVET, an end-to-end training framework that incorporates
an idempotency constraint into the latent representations of a conditional
voice editing model. Experiments on the GLOBE and EARS datasets show that
idempotent training improves identity preservation and editing success rates
while maintaining comparable perceptual quality to the baseline. These results
suggest that idempotency can serve as a practical regularization mechanism
when attribute labels are imperfect or noisy. This work represents an initial step toward understanding how idempotent
objectives improve robustness in conditional generative models. Overall,
idempotency offers a promising direction for improving stability and
robustness in voice editing systems. Future work will explore applying
idempotent objectives to additional attributes beyond age and gender and
studying their effects in both supervised and unsupervised editing settings.

% We release the RIVET framework to support reproducibility and future research.

\section{Generative AI Use Disclosure}

Large language model (LLM) tools were used to assist with proofreading and improving the clarity and fluency of the manuscript. All scientific content, including ideas, methods, experimental design, analysis, and results, was developed and verified by the authors. No generative AI tool is listed as a co-author, and the authors take full responsibility for the contents of this paper.

% \section{Acknowledgments}

% {\color{blue}Acknowledgments should be included only in the camera-ready version, not in the version submitted for review. For regular papers, pages 5 and 6, and for long papers, pages 9 and 10, are reserved exclusively for acknowledgments, disclosures of the use of generative AI tools, and references. No other content may appear on these pages. Any appendices must be contained within the first four pages for regular papers and within the first eight pages for long papers.

% Acknowledgments and references may begin on an earlier page if space permits.}

% \ifcameraready
%      The Interspeech 2026 organizers
% \else
%      The authors
% \fi
% would like to thank ISCA and the organizing committees of past Interspeech conferences for their help and for kindly providing the previous version of this template.

% {\color{blue}
% \section{Generative AI Use Disclosure}
% The extent of Generative AI use must be disclosed. This section may be in the 5th or 6th pages of regular papers, or the 9th or 10th pages of long papers.  ISCA policy says: \textit{All (co-)authors must be responsible and accountable for the work and content of the paper, and they must consent to its submission. Any generative AI tools cannot be a co-author of the paper. They can be used for editing and polishing manuscripts, but should not be used for producing a significant part of the manuscript}.}

% \clearpage

\bibliographystyle{IEEEtran}
\bibliography{mybib}

\end{document}